\documentclass[aps,twocolumn,pra, superscriptaddress]{revtex4-2}
\usepackage{braket}
\usepackage{graphicx}
\usepackage{braket}
\usepackage{amsmath}
\usepackage{stmaryrd}
\usepackage{tabularx,multirow}
\usepackage{dcolumn}
\usepackage{bm}
\usepackage{amsthm}
\usepackage[dvipsnames]{xcolor} 
\usepackage[normalem]{ulem}
\usepackage{hyperref}
\hypersetup{
    colorlinks=true,
    linkcolor=blue,
    filecolor=blue,      
    urlcolor=blue,
    pdftitle={Overleaf Example},
    pdfpagemode=FullScreen,
    citecolor = blue
    }
\usepackage{orcidlink}

\usepackage[toc,page]{appendix}

\begin{document}
\preprint{APS/123-QED}

\title{Optimal quantum spectroscopy using single-photon pulses}

\author{Sourav Das\,\orcidlink{0000-0002-6087-846X}}
\thanks{sourav.iisermohali@gmail.com}
\affiliation{Department of Physics, University of Warwick, Coventry CV4 7AL, United Kingdom}

\author{Aiman Khan\,\orcidlink{0009-0006-0697-8343}}
\thanks{aimankhan1509@protonmail.com}
\affiliation{Manufacturing Metrology Team, Faculty of Engineering, University of Nottingham, Nottingham NG7 2RD, United Kingdom}

\author{Francesco Albarelli\,\orcidlink{0000-0001-5775-168X}}
\thanks{francesco.albarelli@gmail.com}
\affiliation{Università di Parma, Dipartimento di Scienze Matematiche, Fisiche e Informatiche, I-43124 Parma, Italy}

\author{Animesh Datta\,\orcidlink{0000-0003-4021-4655}}
\thanks{animesh.datta@warwick.ac.uk}
\affiliation{Department of Physics, University of Warwick, Coventry CV4 7AL, United Kingdom}

\date{\today}
\begin{abstract}
We provide the ultimate precision attainable in spectroscopy of a quantum emitter using single-photon pulses. We find the maximum for estimating the linewidth to be independent of the details of the emitter's bare Hamiltonian while that for the detunings not to be so. We also identify optimal pulse shapes attaining these precisions.
\end{abstract}

\maketitle
Spectroscopy seeks to estimate parameters  that model a matter system by using light as a probe. Recent studies have used quantum states of light such as single-photons \cite{PhysRevLett.108.093601,chan2018single,li2023single}, squeezed \cite{polzik1992spectroscopy,dorfman2021multidimensional} or entangled states of light \cite{kalachev2007biphoton,PhysRevX.4.011049}, to achieve enhanced precision in spectroscopy beyond what is possible with classical light \cite{dorfman2016nonlinear,mukamel2020roadmap}. However, the maximum possible precision attainable in spectroscopy using quantum states of light remains unidentified.

\begin{figure}[ht]
    \centering
    \includegraphics[width=1\linewidth]{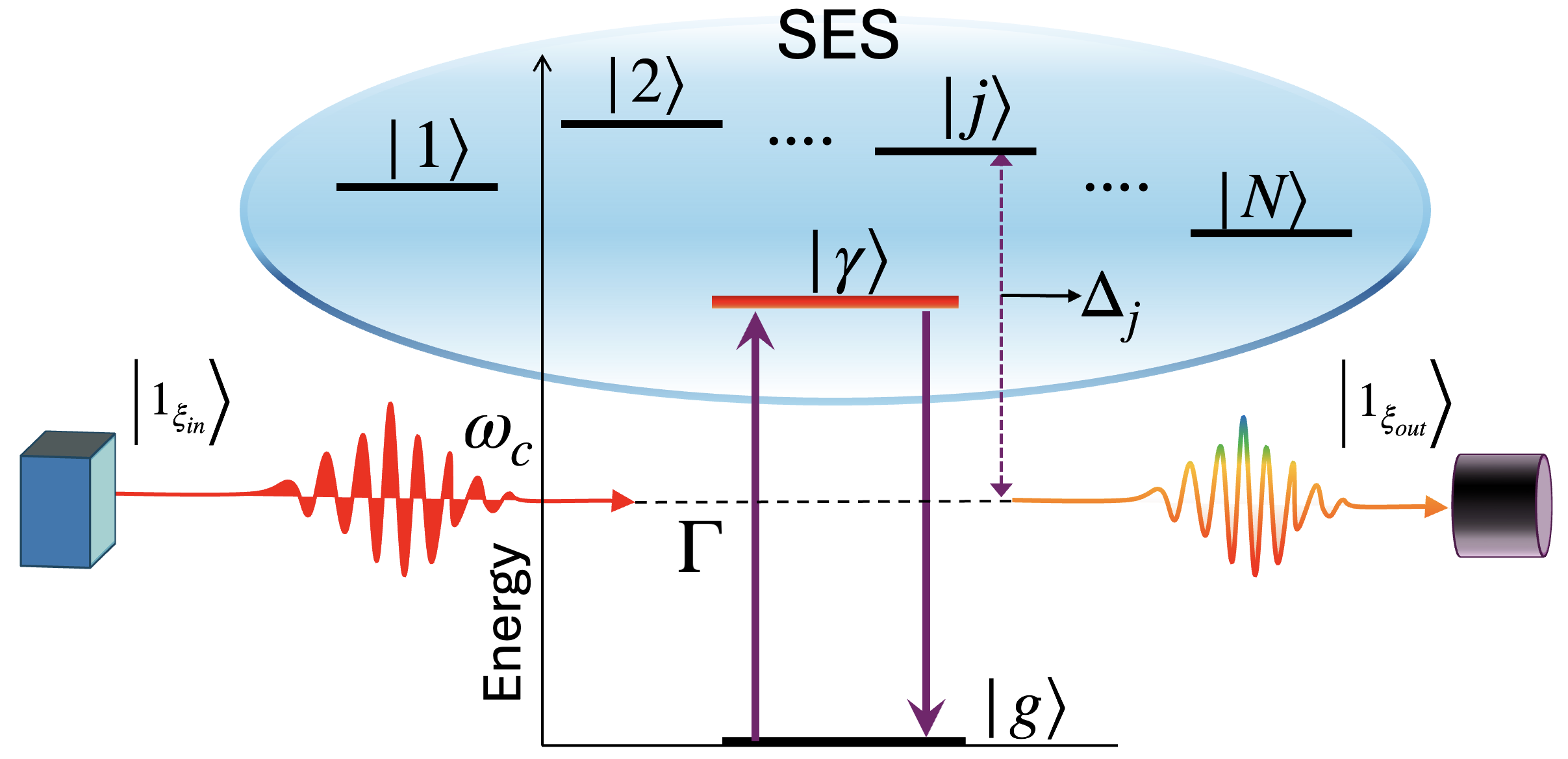}
    \caption{
    The single-photon  $\ket{1_{\xi_\text{in}}}$ excites the emitter from the ground state $\ket{g}$ to the single-excitation subspace (SES) of the emitter which is illustrated by the light blue shaded oval.
    Its elements $\{\ket{j}\}$ $(j=1,2,..,N)$ (with energies $\omega_j$ relative to $\ket{g}$ which is defined to be zero) are illustrated with black solid lines in the SES. The detuning of the $j$-th singly-excited state is $\Delta_j = \omega_j - \omega_c$, where
    the thin horizontal dashed line along the path of the pulse indicates the carrier frequency $\omega_c$.
    $\Gamma$ denotes the coupling strength with the pulse and $\ket{\gamma}$ (thick red line) is a normalized superposition of the states $\{\ket{j}\}$ in the SES. The coupling encodes information of the emitter in the state of the scattered pulse $\ket{1_{\xi_\text{out}}}$.} 
    \label{fig1}
\end{figure}

In this Letter, we provide the ultimate precisions in estimating the 
transition energies (detunings) and linewidths of a quantum emitter starting in its ground state and probed by single-photon pulses \cite{PhysRevLett.100.093603,Stobińska_2009,PhysRevA.83.063842,PhysRevLett.108.093601,chan2018single,li_single-photon_2023}, as illustrated in Fig. \ref{fig1}. 
For conceptual clarity, we assume that all photons emitted from the matter system are detected without loss. This could be physically realized using quantum emitters embedded in wave-guides~\cite{PhysRevLett.113.093603,javadi_numerical_2018,daveau2017efficient}.

Our results are summarised in Table~\ref{tab1}. They are obtained using the tools of quantum estimation theory \cite{paris2009quantum}, which quantifies the ultimate precision with which a parameter can be estimated using quantum probes. 
An upper bound on this precision -- the variance of an unbiased estimator $\hat{\theta}$ of a parameter $\theta$ is set by the Cramér-Rao bound~\cite{paris2009quantum}
\begin{eqnarray}
    \text{Var}(\hat{\theta}) \geq \frac{1}{\mathcal{Q}(\rho_\theta,\theta)}
\end{eqnarray}
where $\mathcal{Q}(\rho_\theta,\theta)$ is the quantum Fisher information (QFI). It depends on the quantum state of the probe $\rho_\theta$ and its derivative with respect to $\theta$.

In spectroscopy, the QFI serves as a measure of the information about $\theta$ contained in the quantum state of the light after it has interacted with the matter system \cite{albarelli2023fundamental,khan2024does,darsheshdar2024role,khan2025tensor}. 
We thus maximize its QFI over all possible input states of light.
 
\textit{Model}---We model the quantum matter system (M) as a point emitter, localised at $z=0$ and possessing a ground state $\ket{g}$ defined to have zero energy. Since we are interested in the single-photon driving of the emitter which is assumed to be in the ground state at the start of the experiment, it will at most accommodate a single-excitation through the course of the interaction. Thus, the total state-space of the emitter can be effectively described by an $(N+1)$-dimensional Hilbert space, where $N$ is the dimension of the singly-excited subspace (SES) of the emitter (see Fig. \ref{fig1}). For example, in a coupled dimer system built from two two-level systems, $\ket{g}$ represents the state with both TLSs in the ground state, and the SES is spanned by $N=2$ states.

The light pulse is a propagating quantized radiation field characterized by a set of frequency-dependent bosonic operators $ a(\omega) $ obeying the commutation relation $ [a(\omega), a^\dagger(\omega')] = \delta(\omega - \omega') $. The free field Hamiltonian is given by $ H_\text{P} = \int d\omega \, \omega a^\dagger(\omega) a(\omega) $ (assuming natural units characterised by $\hbar=1$). We also assume that the incoming quantum optical pulse is sufficently narrow in spectral bandwidth $B\ll\omega_c$ around the carrier frequency $\omega_c$, a reasonable assumption for optical frequencies. 

The electromagnetic field then couples to the emitter through the dipole interaction Hamiltonian $H_\text{MP}=-\vec{\textbf{d}}\cdot \vec{\textbf{E}}$, where $\vec{\textbf{d}}$ is the emitter dipole moment, and
\begin{equation}
\label{eq:E}
    \vec{\textbf{E}} = i\hat{\textbf{e}}\int_{0}^{\infty} d\omega \sqrt{\frac{\omega}{4\pi\epsilon_0 c A}}\,a(\omega) e^{-i\omega t} + \text{h.c.}
\end{equation}
is the linearly polarised electric field operator in the paraxial regime~\cite{deutsch1991paraxial}, $A$ being the transverse quantisation area, and $\hat{\textbf{e}}$ the unit polarisation vector. Assuming also that the incoming light pulse couples only very weakly to the matter system $M$~(so timescales for light-matter dynamics are well separated from matter-only dynamics), we can employ the rotating wave and Markovian white noise approximations \cite{scully1997quantum, loudon2000quantum,baragiola2014open} to express the total interaction-picture light-matter Hamiltonian as (see Appendix \ref{appendix1} for a microscopic derivation):
\begin{equation} 
\label{eq1}
    H(t) = H_\text{M} - i\sqrt{\Gamma} \left( \Sigma^\dagger a(t) - \Sigma a^\dagger(t) \right),
\end{equation}
where $ H_\text{M}$ is the time-independent Hamiltonian of the emitter M, and $\Gamma$ is the Weisskopf-Wigner rate of spontaneous emission characterising the strength of the light-matter coupling, inversely proportional to the emitter's radiative lifetime which in turn determines the natural linewidth of the emitted light $(\Gamma/2\pi)$ to be estimated later. The time-dependent field operators $ a(t) = \int_{-\infty}^{\infty} d\omega \, a(\omega) e^{-i(\omega - \omega_c)t} / \sqrt{2\pi} $ are the Fourier transform of the field operators $a(\omega)$, dubbed the ``quantum white-noise operators", which in the Markovian limit satisfy the commutation relation $ [a(t), a^\dagger(t')] = \delta(t - t') $. 

The transition operator that couples to the field is taken to be $ \Sigma = \ket{g}\bra{\gamma}$, where $\ket{\gamma}$ is a normalized vector in the SES that typically depends on the dipole moment operator of the emitter~\cite{chan2018single,ko2022dynamics}. This form of the transition operator that connects only the ground state to the SES states and excludes all higher excited states is employed under the constraint of the traveling light pulse carrying a single-excitation to the ground state of the emitter at the starting time $t=t_0$ before the pulse interacts with M (see Appendix \ref{appendix1}).

The light-matter Hamiltonian $H(t)$ preserves the total number of excitations which is defined by the operator $\textbf{N}_{\text{MP}}=\mathbf{I}_{\text{M}}+\int_{0}^{\infty} d\omega a^\dagger(\omega)a(\omega)$, where $\mathbf{I}_\text{M}$ is the identity operator in the SES. 
At $t = t_0$, the joint light-emitter state is $\ket{\Psi(t_0)}=\ket{g}\ket{1_{\xi_\text{in}}}$ where the incident single-photon state of the field is given by,
\begin{eqnarray}
    \ket{1_{\xi_\text{in}}}= \int_{-\infty}^{\infty} d\tau \,\xi_\text{in}(\tau)a^\dagger(\tau)\ket{0_\text{P}},
\end{eqnarray}
where $a^\dagger(\tau)\ket{0_\text{P}}$ is an element of the temporal mode basis of the single-photon and $\xi_\text{in}(\tau)$ is the incident wave packet of the pulse. This single-photon excites the emitter from $\ket{g}$ to the SES which subsequently spontaneously decays into $\ket{g}$ by emitting a photon, assuming M couples to no other environments. Thus, the final light-emitter state 
at times $t\gg 1/\Gamma$ takes the form $\ket{\Psi(t\gg 1/\Gamma)}=
\ket{g}\ket{1_{\xi_\text{out}}}$, where $\ket{1_{\xi_\text{out}}}$ is the scattered single-photon state. Solving the Schr\"{o}dinger equation corresponding to light-matter coupling in Eq.~(\ref{eq1}), we obtain~\cite{PhysRevA.93.063807,ko2022dynamics}
\begin{eqnarray}
\label{eq2}
    \xi_{\text{out}}(\tau) = \xi_{\text{in}}(\tau) - \int_{-\infty}^{\infty} dt^\prime\, \xi_{\text{in}}(t^\prime) f_{\text{M}}(\tau-t^\prime),\\
    \text{with }f_{\text{M}}(t)= \Gamma\Theta(t)\bra{\gamma} \exp{\left(-iH_\text{M}-\Gamma\frac{\ket{\gamma}\bra{\gamma}}{2}\right)t} &\ket{\gamma},\nonumber
\end{eqnarray}
where $\Theta(\cdot)$ is the Heaviside theta function. In the frequency domain
\begin{eqnarray}
\label{eq10}
    \tilde{\xi}_{\text{out}}(\omega) = \tilde{\xi}_{\text{in}}(\omega) \left(1-\tilde{f}_{\text{M}}(\omega)\right),
\end{eqnarray}
where $\tilde{\xi}_{\text{in/out}}(\omega)=\int_{-\infty}^{\infty} d\tau \xi_{\text{in/out}}(\tau) e^{i\omega t}/\sqrt{2\pi}$ and $\tilde{f}_{\text{M}}(\omega)=\int_{-\infty}^{\infty} d\tau f_{\text{M}}(\tau) e^{i\omega t}$ is the Fourier transform of the time-domain emitter kernel $f_{\text{M}}(t)$. Indeed,
\begin{eqnarray}
\label{matterresponse}
     \tilde{f}_{\text{M}}(\omega) = \frac{2i\chi_\text{M}(\omega)}{1+i\chi_\text{M}(\omega)},\\
    \text{with } \chi_\text{M}(\omega) = \Gamma\bra{\gamma}\llbracket(\omega\mathbf{I}_\text{M}-H_\text{M})&\rrbracket^{-1}\ket{\gamma}/2, \nonumber
\end{eqnarray}
where $\llbracket.\rrbracket^{-1}$ denotes the operator inverse.

Since the final state of the scattered pulse is a pure state $\ket{1_{\xi_{\text{out}}}}$, we can treat the total evolution of the pulse as a unitary encoding process with an effective unitary operator $U_\text{B}$ such that $\ket{1_{\xi_{\text{out}}}} = U_\text{B}\ket{1_{\xi_{\text{in}}}}$. Using Eq. (\ref{eq10}),
\begin{eqnarray}
\label{blackboxunitary}
    U_\text{B} = \int_{-\infty}^{\infty} d\omega \left(\frac{1-i\chi_\text{M}(\omega)}{1+i\chi_\text{M}(\omega)}\right) \ket{1_\omega}\bra{1_\omega},
\end{eqnarray}
where $\ket{1_\omega}=a^\dagger(\omega)\ket{0_\text{P}}$.
This operator is diagonal in the frequency basis, and its unitarity is ensured by the diagonal elements which represent a frequency dependent phase.
Therefore, we can define a dimensionless effective Hamiltonian that generates the unitary $U_\text{B} = \text{exp}(-i H_\text{B})$, expressed explicitly as
\begin{eqnarray}
\label{eq14}
     H_\text{B} = \int_{-\infty}^{\infty} d\omega\, 2\,\text{tan}^{-1}\left(\chi_\text{M}(\omega)\right)\ket{1_\omega}\bra{1_\omega}.
\end{eqnarray}
This effective Hamiltonian formulation of the unitary parameter encoding onto the single-photon pulse due to the light-matter interaction forms the basis of the evaluation of quantum precision bounds, and is one of our central results.

\textit{Spectroscopy}---For unitary parameter encoding generated by the Hamiltonian operator $H_\text{B}$, the following tight upper bound to the QFI of estimating a parameter $\theta$ encoded in the eigenvalues of $H_\text{B}$ can be obtained \cite{PhysRevA.90.022117,pang2017optimal}, 
\begin{eqnarray}
\label{eqq9}
    \mathcal{Q}(\ket{1_{\xi_{\text{out}}}},\theta) \leq (\mu^{\theta}_{\text{max}}-\mu^{\theta}_{\text{min}})^2 ~
    \equiv \mathcal{Q}_{\text{max}}(\theta).
\end{eqnarray}
Here $\mu^{\theta}_\text{min/max}$ represent the minimum/maximum eigenvalues of the operator $\partial_\theta H_\text{B}.$
These correspond to the global minima and maxima of the function  
$$\mathcal{X}_\text{M}^\theta(\omega)=\frac{2\partial_\theta\chi_\text{M}(\omega)}{1+\chi_\text{M}(\omega)^2}.
$$
The upper bound in Eq.~\eqref{eqq9} is saturated when the incident pulse is an equally weighted superposition of the eigenstates of $\partial_\theta H_\text{B}$ corresponding to $\mu^{\theta}_{\text{max}}$ and $\mu^{\theta}_{\text{min}}$. In our case, this state is given by 
\begin{eqnarray}
\label{eqopt}
    \ket{1_\text{opt}} = \frac{1}{\sqrt{2}}\left(\ket{1_{\Omega^{\theta}_{\text{max}}}}+e^{i\varphi}\ket{1_{\Omega^{\theta}_{\text{min}}}}\right),
\end{eqnarray}
wherein the frequencies are given by the solutions of 
$\mathcal{X}_\text{M}^\theta(\Omega^{\theta}_{\text{max/min}}) = \mu_{\text{max/min}}$, where $e^{i\varphi}$ is an arbitrary phase.
For a degenerate $\partial_\theta H_\text{B}$ with non-unique states $\ket{1_{\Omega^{\theta}_{\text{max/min}}}}$, any such superposition will saturate the bound.
In the continuous frequency space, these states are represented by a combination of Dirac delta functions with peaks located at the frequencies $\Omega^{\theta}_{\text{max/min}}$. 

We now obtain the  ultimate precision of estimating the linewidth $\Gamma$ of the single-photon transition, and the detunings $\Delta_j$ in the SES of the emitter.

\textit{$\Gamma$ estimation}---Using Eq.~\eqref{matterresponse} and that $H_{\text{M}}$ is independent of $\Gamma,$ $\partial_\Gamma\chi_\text{M}(\omega)=\chi_\text{M}(\omega)/\Gamma$.
Noting that $-1\leq\frac{2\chi_\text{M}(\omega)}{1+\chi_\text{M}(\omega)^2}\leq 1$, where the equalities hold when $\chi_\text{M}(\omega)=\pm1$, the maximum and the minimum values of $\mathcal{X}_\text{M}^\Gamma(\omega)$ are $\pm1/\Gamma.$
This leads to
\begin{eqnarray}
\label{eqgamma}
    \mathcal{Q}(1_{\ket{\xi_{\text{out}}}},\Gamma)\leq \left(2/\Gamma\right)^2.
\end{eqnarray}
Curiously, this upper bound has no explicit dependence on the emitter Hamiltonian $H_\text{M}$ or the SES vector $\ket{\gamma}$, allowing us to conclude that quantum mechanical bound on precision of lifetime estimation only depends on the level structure of $H_\text{M}$ insofar as its effect on the Weisskopf-Wigner rate $\Gamma$ corresponding to the emitter.

The optimal pulse that saturates this bound is an equally weighted superposition of two frequency-mode states with frequencies $\Omega^{\Gamma}_{\text{max/min}}$ which satisfy $\chi_\text{M}(\omega)=\pm 1$.
These equations have real solutions in $\omega$ 
for a general emitter system, ensuring the bound is tight (see Appendix \ref{appendix2} for details). 

\begin{figure}[ht]
    \centering
    \includegraphics[width=1\linewidth]{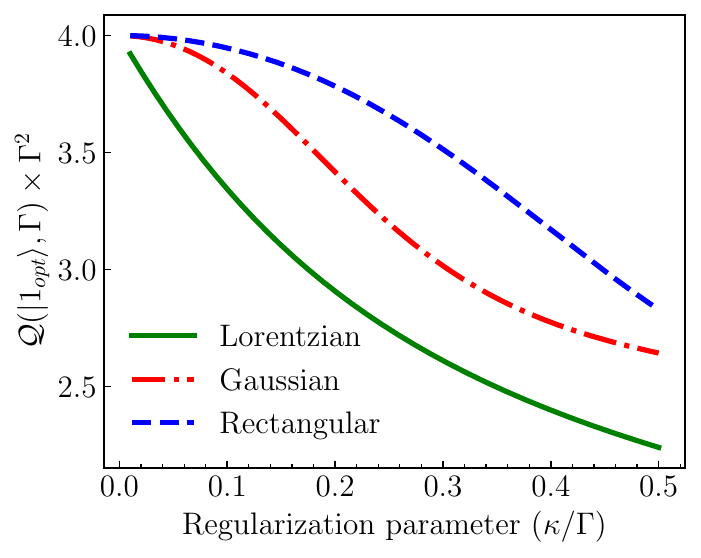}
    \caption{The dimensionless QFI of estimating $\Gamma$ for a TLS using regularized optimal incident pulses as a function of the regularization parameter $\kappa/\Gamma$. The regularized delta functions are defined as the Lorentzian $\delta_\kappa(\omega)=\kappa/(\pi(\kappa^2+\omega^2))$, the Gaussian $\delta_\kappa(\omega)=e^{-\frac{\omega^2}{2\kappa^2}}/(\kappa\sqrt{2\pi})$, and, the rectangular $\delta_\kappa(\omega)=\Theta(\omega)\Theta(\kappa-\omega)/\kappa$.}
    \label{fig2}
\end{figure}

For the specific case of a two-level system (TLS), these solutions are given by: $\Omega^{\Gamma}_{\text{max/min}}=\Delta\pm\Gamma/2$, where $\Delta=\omega_0-\omega_c$ is the detuning and $\omega_0$ is the transition frequency of the TLS. According to Eq. (\ref{eqopt}) the optimal incident pulse for a TLS is
\begin{eqnarray}
    \ket{1_{\text{opt}}}=\frac{1}{\sqrt{2}}\left[\ket{1_{\Delta+\frac{\Gamma}{2}}}+e^{i\varphi}\ket{1_{\Delta-\frac{\Gamma}{2}}}\right].
\end{eqnarray}
This optimal incident pulse is expressed in the frequency domain by a combination of two delta functions $\vert\tilde{\xi}_{\text{opt}}(\omega)\vert^2 = [\delta(\omega-\Delta+\Gamma/2)+\delta(\omega-\Delta-\Gamma/2)]/2$. Engineering such a pulse-shape would be experimentally challenging due to the singular nature of the spectral profile.
However, the ultimate precision can be approached with a regularized spectral profile of the pulse. In Fig. (\ref{fig2}) we plot the QFI for the optimal incident pulse constructed by three distinct regularized representations of the Dirac delta $\delta_{\kappa}(\omega)$
where $\delta(\omega) = \underset{\kappa\rightarrow 0}{\lim}\,\delta_\kappa(\omega)$ as a function of the dimensionless regularization parameter $\kappa/\Gamma$. For all these representations, the QFI monotonically approaches the upper-bound in the limit $\kappa\ll\Gamma$.

\textit{Detuning estimation}---Spectroscopy also involves estimating 
the tranistion energies of the emitter Hamiltonian $H_\text{M}$. We now present ultimate precisions for estimating these detunigs $\Delta_j=\bra{j}H_\text{M}\ket{j}=\omega_j-\omega_c$~($\omega_j$ is the energy of the $j$th state), defined as the diagonal elements of $H_\text{M}$ in the SES basis $\{\ket{j}\}$. See Fig. \ref{fig1}.

Differentiating $\chi_\text{M}(\omega)$ from Eq. (\ref{matterresponse}) with respect to $\Delta_j$ we get $\partial_{\Delta_j} \chi_\text{M}(\omega) = \Gamma |\bra{j}\llbracket\omega\mathbf{I}_\text{M}-H_\text{M}\rrbracket^{-1}\ket{\gamma}|^2/2$. This indicates that the function $\mathcal{X}_\text{M}^{\Delta_j}(\omega)$ is non-negative and it approaches the minima zero for $\omega \ll \epsilon_{\min}$ or $\omega \gg \epsilon_{\max}$, where $\epsilon_{\min/\max}$ are the minimum and maximum eigenvalues of $H_\text{M}$. The global maxima of $\mathcal{X}_\text{M}^{\Delta_j}(\omega)$, denoted as $\mu^{\Delta_j}_{\max}$, depends on the emitter parameters and can be calculated for any emitter system.
This leads to the following precision bound: $\mathcal{Q}(|1_{\xi_{\text{out}}}\rangle,\Delta_j) \leq \left(\mu^{\Delta_j}_{\max}\right)^2$.

For a TLS, this upper bound simplifies to $\mathcal{Q}(|1_{\xi_{\text{out}}}\rangle,\Delta) \leq (4/\Gamma)^2$.
The corresponding optimal pulse is given by $|1_{\text{opt}}\rangle = \frac{1}{\sqrt{2}} [|1_{\Delta}\rangle + e^{i\varphi} |1_{\infty}\rangle]$, where $|1_{\Delta}\rangle$ is a frequency-mode state resonant with the detuning $\Delta$, and $|1_{\infty}\rangle$ is a frequency-mode state far from resonance.

\begin{table*}[ht]
\begin{center}
\begin{tabular}{c|c|c}
\multirow{3}{*}{} 
& $N+1$ level emitter & TLS \\
\hline
\hline
\multirow{3}{*}{$\mathcal{Q}_{\text{max}}(\Gamma)$} & & \\
& $\left(\dfrac{2}{\Gamma}\right)^2$ & $\left(\dfrac{2}{\Gamma}\right)^2$ \\

\multirow{3}{*}{$\ket{1_{\text{opt}}}$} & & \\
& $\dfrac{1}{\sqrt{2}}\left[\ket{1_{\Omega^{\Gamma}_{\text{max}}}}+e^{i\varphi}\ket{1_{\Omega^{\Gamma}_{\text{min}}}}\right]$ & $\dfrac{1}{\sqrt{2}}\left[\ket{1_{\Delta+\frac{\Gamma}{2}}}+e^{i\varphi}\ket{1_{\Delta-\frac{\Gamma}{2}}}\right]$ \\
& & \\
\hline
\multirow{3}{*}{$ \mathcal{Q}_{\text{max}}(\Delta_j)$ } & & \\
& $\left(\mu^{\Delta_j}_{\max}\right)^2 = \left( \max \mathcal{X}_\text{M}^{\Delta_j}(\omega) \right)^2$ & $\left(\dfrac{4}{\Gamma}\right)^2$ \\

\multirow{3}{*}{$\ket{1_{\text{opt}}}$} & & \\
& $\dfrac{1}{\sqrt{2}} \left[|1_{\Omega^{\Delta_j}_{\text{max}}}\rangle + e^{i\varphi} |1_{\infty}\rangle\right]$ & $\dfrac{1}{\sqrt{2}} \left[|1_{\Delta}\rangle + e^{i\varphi} |1_{\infty}\rangle\right]$ \\ && \\
\hline
\end{tabular}
\caption{ Upper bounds to the QFI ($\mathcal{Q}_{\text{max}}(\cdot)$) and the optimal incident pulse states ($\ket{1_{\text{opt}}}$) for estimating the linewidth $\Gamma$ and detunings $\Delta_j$ with single-photons for a $N+1$ level emitter (2nd column)  and a two-level emitter (3rd column).}
\label{tab1}
\end{center}
\end{table*}

\textit{Conclusions}---We have identified the ultimate precision of estimating two key spectroscopic parameters as summarised in Table. \ref{tab1}. The corresponding optimal incident pulses are shown to be combinations of delta functions in the frequency domain, whose locations are specific to the emitter system and the estimation task.  

Our theoretical limit of $4/\Gamma^2$ for $\Gamma$ estimation surpasses the maximum precision of about $2.5/\Gamma^2$ achieved for a TLS with both Gaussian and rectangular input pulses, and the maximum of $2/\Gamma^2$ achieved for decaying and rising exponential pulses \cite[Fig. 2]{albarelli2023fundamental}. Indeed, the pulse which is most efficiently absorbed (rising exponential) is not the most informative spectroscopically~\cite{stobinska2009perfect,PhysRevA.96.033817,PhysRevLett.111.103001,albarelli2023fundamental}.

A recent study with pulsed coherent light probing a TLS reveal that the QFI per photon respects the same bound \cite{karthik} of $4/\Gamma^2$ for estimating $\Gamma$. This suggests that our bounds are valid beyond single-photon pulses. Future research may investigate the validity of our bounds for other quantum probe states and more general emitter states, as well as for imperfect couplings of the traveling light state to the matter system.

We thank Elnaz Darsheshdar, Karthik Chinni and Nicolás Quesada for fruitful discussions.
This work has been funded, in part, by an EPSRC New Horizons grant (EP/V04818X/1) and the UKRI (Reference Number: 10038209) under the UK Government’s Horizon Europe Guarantee for the Research and Innovation Programme under agreement 101070700 (MIRAQLS). Computing facilities were provided by the Scientific Computing Research Technology Platform of the University of Warwick.

\bibliographystyle{apsrev4-2}
\bibliography{bib}
\appendix
\section{Microscopic derivation of the light-matter Hamiltonian}
\label{appendix1}
In this section, we derive the total light-matter Hamiltonian in Eq. (\ref{eq1}) from the dipole interaction Hamiltonian $H_\text{MP}=-\vec{\textbf{d}}\cdot \vec{\textbf{E}}$. The dipole operator of our emitter system takes the following form:
\begin{eqnarray}
    \vec{\textbf{d}} = \sum_{k=1}^{N} \vec{d}_{gk} \ket{k}\bra{g} \,+\, \text{h.c.} ,
\end{eqnarray}
where $\vec{d}_{gk}=\bra{g}\vec{\textbf{d}}\ket{k}$ are the relevant matrix elements of the dipole operator, and $\{\ket{k}\}$ is a basis of the SES. Note that this dipole operator neglects the matrix elements within the SES. This is physically valid because the energy separation within the singly excited states is always lower than the energy separations between the ground state and the excited states with which the optical field is resonant. However, the transitions within the excited states can happen due to the dipole-dipole coupling within the molecule which are accounted for by the bare emitter Hamiltonian $\bar{H}_\text{M}$. Since we assume the ground state has zero energy, $\bar{H}_\text{M}$ has support only in the SES.  Thus, the total light-matter Hamiltonian can be written as $H = \bar{H}_\text{M}+H_\text{P}+H_\text{MP}$. 

Transforming to the interaction picture with respect to the static Hamiltonian $H_0 = \omega_c\sum_{k=1}^{N}\ket{k}\bra{k} + H_\text{P}$ and rotating away the rapidly
oscillating terms, the total light-matter Hamiltonian takes the following form
\begin{eqnarray}
    H(t) = H_\text{M} \\
    -i\Bigg[\sum_{k=1}^{N}D_k &\ket{g}\bra{k}\otimes \int_{0}^{\infty} d\omega \, \mathcal{E}(\omega)a(\omega)e^{-i(\omega-\omega_c)t}-\text{h.c.}\Bigg],\nonumber
\end{eqnarray}
where $D_k = \vec{d}_{gk}\cdot\hat{\textbf{e}}$, and, the amplitude of the electric field $\mathcal{E}(\omega)=\sqrt{\frac{\omega}{4\pi\epsilon_0 c A}}$.  $H_\text{M}=\bar{H}_\text{M}-\omega_c\sum_{k=1}^{N}\ket{k}\bra{k}$ is the detuned emitter Hamiltonian which is referred to as the emitter Hamiltonian in the main text for brevity.

Now, we impose the white-noise approximation which is often referred to as the Markovian approximation in quantum optics, which assumes that the incident pulse is sufficiently narrow-band around the carrier frequency $\omega_c$. Under this approximation, the amplitude of the electric field operator becomes only dependent on $\omega_c$ and the limits to the integral over the frequency modes can be set to $-\infty$ to $+\infty$. Thus, we can write,
\begin{eqnarray}
    H(t) = H_\text{M} - i\sqrt{\Gamma} \left( \Sigma^\dagger a(t) - \Sigma a^\dagger(t) \right),
\end{eqnarray}
where the coupling strength $$\Gamma = \left(\frac{\omega_c}{4\pi\epsilon_0 c A}\right)\times\sum_{k=1}^{N}\vert D_k\vert^2$$, and $\Sigma = \ket{\gamma}\bra{g}$ where $\ket{\gamma}$ is a normalized state in the SES expressed as $\ket{\gamma}\propto \sum_{k=1}^{N} D_k\ket{k}$ where the normalization factor is absorbed in $\Gamma$.

\section{Proof of existence of solutions to $\chi_\text{M}(\omega)=\pm1$}
\label{appendix2}
Writing the emitter Hamiltonian in spectral decomposition $H_\text{M}=\sum_{k=1}^{N} \epsilon_k \ket{\epsilon_k}\bra{\epsilon_k}$, where $N$ is the dimension of the singly excited subspace of the emitter, we can get the following expression:
$\chi_\text{M}(\omega)=\frac{\Gamma}{2}\sum_{k=1}^{N} \frac{\vert\langle\gamma\vert\epsilon_k\rangle\vert^2}{\omega-\epsilon_k}.
$
See $\chi_\text{M}(\omega)$ has poles at the frequencies $\omega=\epsilon_k$. Thus, $\chi_\text{M}(\omega)\rightarrow +\infty$ when $\omega\rightarrow\epsilon_k^+$ and $\chi_\text{M}(\omega)\rightarrow -\infty$ when $\omega\rightarrow\epsilon_k^-$ for all $k$. Therefore, $\chi_\text{M}(\omega)$ is continuous in the interval $(\epsilon_k,\epsilon_{k+1})$ and has a range of $(-\infty,+\infty)$. Thus, applying the intermediate value theorem we can state $\chi_\text{M}(\omega)$ has at least 1 real solution to $\chi_\text{M}(\omega)=c$ for any real number $c$. This proves that the equations $\chi_\text{M}(\omega)=\pm 1$ have real solutions in $\omega$.

\end{document}